\documentclass[conference]{IEEEtran}
\IEEEoverridecommandlockouts
\usepackage{amsmath}
\usepackage{amssymb}
\usepackage{graphicx}
\usepackage{algorithm}
\usepackage{algorithmicx}
\usepackage{algpseudocode}
\usepackage{float}
\usepackage{titlesec}
\usepackage{booktabs}
\usepackage{multirow}
\usepackage{newclude}

\usepackage[paperwidth=8.5in, paperheight=11in]{geometry}
\titlespacing{\section}{0pt}{3.5ex plus 1ex minus .2ex}{2.3ex plus .2ex}
\titlespacing{\subsection}{0pt}{3.25ex plus 1ex minus .2ex}{1.0ex plus .2ex}
\begin{document}

\title{Personalized Federated Deep Reinforcement Learning-based Trajectory Optimization for Multi-UAV Assisted Edge Computing
\thanks{This work is partially supported by the Youth Foundation Project of Zhejiang Lab (No. K2023PD0AA01), partially supported by the Research Initiation Project of Zhejiang Lab (No. 2022PD0AC02), partially supported by the National Natural Science Foundation of China under Grant No. 62002170, partially supported by Zhejiang Lab Open Research Project (No. K2022PD0AB05). Corresponding author: Chuan Ma.}
}

\author{
\IEEEauthorblockN{
Zhengrong Song\IEEEauthorrefmark{1},
Chuan Ma\IEEEauthorrefmark{2},
Ming Ding\IEEEauthorrefmark{3},
Howard H. Yang\IEEEauthorrefmark{4},
Yuwen Qian\IEEEauthorrefmark{1},
Xiangwei Zhou\IEEEauthorrefmark{5}}
\IEEEauthorblockA{\IEEEauthorrefmark{1}Nanjing University of Science and Technology, Nanjing, China}
\IEEEauthorblockA{\IEEEauthorrefmark{2}Zhejiang Lab, Hangzhou, China \;
\IEEEauthorrefmark{4}ZJU-UIUC Institute, Zhejiang University, Haining 314400, China}
\IEEEauthorblockA{\IEEEauthorrefmark{3}Data61, CSIRO, Sydney, Australia \; 
\IEEEauthorrefmark{5}Louisiana State University, Baton Rouge, LA 70803, USA}
Email: zhengrongsong@njust.edu.cn, chuan.ma@zhejianglab.edu.cn}
\maketitle

\begin{abstract}
In the era of 5G mobile communication, there has been a significant surge in research focused on unmanned aerial vehicles (UAVs) and mobile edge computing technology. UAVs can serve as intelligent servers in edge computing environments, optimizing their flight trajectories to maximize communication system throughput. Deep reinforcement learning (DRL)-based trajectory optimization algorithms may suffer from poor training performance due to intricate terrain features and inadequate training data. To overcome this limitation, some studies have proposed leveraging federated learning (FL) to mitigate the data isolation problem and expedite convergence. Nevertheless, the efficacy of global FL models can be negatively impacted by the high heterogeneity of local data, which could potentially impede the training process and even compromise the performance of local agents. This work proposes a novel solution to address these challenges, namely personalized federated deep reinforcement learning (PF-DRL), for multi-UAV trajectory optimization. PF-DRL aims to develop individualized models for each agent to address the data scarcity issue and mitigate the negative impact of data heterogeneity. Simulation results demonstrate that the proposed algorithm achieves superior training performance with faster convergence rates, and improves service quality compared to other DRL-based approaches.
\end{abstract}


\section{INTRODUCTION}
The speedy advancement of contemporary communication technology has brought about an intensified demand for computing processing from mobile users. However, conventional cloud computing models fall short of catering to the requirements of massive data processing. In recent times, mobile edge computing has considerably enhanced the efficacy of task processing, providing mobile users with high-quality network services and minimal latency~\cite{corcoran2016mobile}. Nevertheless, certain remote regions present considerable challenges in the deployment of edge servers. In light of their portability, flexibility, and high mobility, UAV-assisted edge computing has emerged as a burgeoning trend in such scenarios~\cite{fotouhi2019survey}.

Trajectory optimization is a vital component in UAV edge computing~\cite{1}. As an action-learn-based algorithm, reinforcement learning (RL) shows its potential advantages in addressing the complex environment~\cite{9727746}. In~\cite{yan2018path}, an improved Q-learning-based algorithm was proposed for path planning in an unknown antagonistic environment. Authors in~\cite{9739975} designed a Monte Carlo Tree Search (MCTS)-based path planning scheme, which can plan the flight path of a UAV in a dynamic environment reasonably. However, the aforementioned algorithms can only be applied to discrete and low-dimensional action spaces~\cite{nguyen2020deep}. In reality, each state of the agent contains numerous action possibilities, and previous work [4-6] may suffer from the curse of high-dimensional action, leading to a slow or even non-convergence rate.

In order to solve the above problems, combining the deep neural network with RL, a deep reinforcement learning (DRL) algorithm is utilized to solve the dimensional disaster of huge states and action spaces~\cite{9465671},~\cite{bayerlein2021multi}. Moreover, a single UAV provides limited services, while multi-UAV can expand the service scope and provide better service quality through mutual cooperation. Thus, authors in~\cite{wang2020multi} proposed a multi-agent deep deterministic policy gradient (MADDPG) based trajectory optimization algorithm to realize the fairness of user service and the server terminal load, as well as minimizing the UAV's energy consumption. In~\cite{zhao2022multi}, a collaborative multi-agent DRL framework was proposed to obtain the joint strategy of trajectory design, task allocation, and power management. In~\cite{gao2021game}, a potential game method was proposed to solve the service allocation problem of multi-UAV in advance and then optimize the trajectory.

Nevertheless, multi-agent deep reinforcement learning faces challenges like low learning efficiency and slow convergence rates in complex dynamic environments. These issues arise due to the fact that agents interact and learn from each other, and changes for different clients can affect each other's decisions, leading to instability. As a new learning paradigm, federated learning (FL) has become more and more popular in recent years. Through FL, distributed learning schemes can be efficiently performed among multiple participants at a low communication cost~\cite{2021A}, and the combination of FL and RL has become a natural solution to address the multi-UAV trajectory optimization. Recently, the authors in~\cite{kwon2020multiagent} proposed a federated multi-agent deep deterministic policy gradient (F-MADDPG), in which model parameters are shared, thus greatly reducing communication delay and overhead. Although this algorithm has a good performance in MEC system allocation scheduling problems, it is not practical that different clients share a single model, especially when clients locate in heterogeneous situations~\cite{ziying2021towards}.

To address the aforementioned issues, in this paper, we propose a UAV trajectory optimization algorithm based on personalized federated deep reinforcement learning (PF-DRL). In detail, each UAV no longer uses a single global model. By improving the aggregation process of FL, the global model and the local model are aggregated with a moderate weight to train a personalized model. Such an approach not only amplifies the learning efficiency but also empowers each UAV to make personalized action decisions guided by the local training model, thus ensuring the overall learning performance. Specifically, the main contributions of this paper are summarized as follows.
\begin{itemize}
  \item We build a multi-UAV-assisted mobile edge computing model for the complex and dynamic environment. While optimizing the flight trajectory of each UAV, the overall energy consumption is reduced as required.
  \item Since each UAV is trained locally based on the MADDPG algorithm. We, for the first time, propose a UAV trajectory optimization method based on personalized federated multi-agent deep deterministic policy gradient (PF-MADDPG).
  \item Finally, the proposed method is synthetically simulated. The results show that the learning efficiency and convergence rate of our algorithm is significantly improved without degrading the overall performance.
\end{itemize}

\section{SYSTEM MODEL AND PROBLEM FORMULATION}
In this section, the system model is presented. As shown in Fig. 1, the UAV serves as a mobile edge computing server to provide computing migration and data storage services for a group of users on the ground. We assume that there are $M$ randomly distributed ground users in a square area with a side length $L^{max}$, and the set of users is denoted as $m \in \mathcal{M}=\{1,2, \ldots, M\}$. We define all of the users served by $N$ UAVs where $n \in \mathcal{N}=\{1,2, \ldots, N\}$, which are set to fly in the designated area.

We divide the flight time of the UAV in an episode into $T$ different time intervals, and a single time slot (TS) $t \in \mathcal{T}=\{1,2, \ldots, T\}$. In a specific TS, all users will move to random locations in the area. The positions of users can be expressed as $\mathcal{U}_{\text {\emph{m} }}(t)=\left[x_{\text {\emph{m} }}(t), y_{\text {\emph{m} }}(t), 0\right]$, and the locations of UAVs are denoted as $\mathcal{P}_{\text {\emph{n} }}(t)=\left[x_{\text {\emph{n} }}(t), y_{\text {\emph{n} }}(t), H\right]$. We consider that each UAV is flying at a fixed altitude $H$. In addition, there are obstacles such as buildings and trees in our environment, therefore each UAV should keep a distance from them, as well as other UAVs, during the execution of tasks.

The horizontal distance between the \emph{m}-th user and the \emph{n}-th UAV in the \emph{t}-th TS can be expressed as
\begin{equation}
d_{m,n}(t)= \sqrt{\left\|x_{\emph{m }}(t)-x_{\emph{n }}(t)\right\|^2+\left\|y_{\emph {m }}(t)-y_{\emph {n }}(t)\right\|^2}.
\end{equation}

Our communication link between the UAV and users is line-of-sight (LoS). The calculation of channel power gain follows the free space path loss model, which can be expressed as
\begin{equation}
g_{m, n}(t)=\frac{\rho_0}{H^2+d_{m,n}^2(t)},
\end{equation}
where $\rho_0$ represents the channel's power gain at the reference distance. Then, the uploading data rate from the \emph{m}-th user to the \emph{n}-th UAV in the \emph{t}-th TS can be expressed as
\begin{equation}
R_{m, n}(t)=B_{m}\log_{2}{(1+\frac{P_{u}g_{m, n}(t)}{\sigma^2})},
\end{equation}
where $B_{m}$ represents the channel bandwidth allocated by the \emph{n}-th UAV to the \emph{m}-th user, so that $\sum_{m=1}^{M^{'}}{B_{m}}=B$ and $M^{'}$ is number of users served by UAV simultaneously. In order to simplify the model, we assume that the UAV shares the channel bandwidth equally with the serving users. In addition, $P_{u}$ denotes the transmission power of the \emph{m}-th user and $\sigma^2$ is the power of additive white Gaussian noise.

We also consider the energy consumption generated by the UAV during task execution, mainly including flight energy consumption and computing energy consumption. We assume that the UAV flies at a constant speed v(t), and in one TS, the energy consumed by \emph{n}-th UAV flight can be defined as
\begin{equation}
E_{n}^f (t)=\kappa v_{n}^2(t),
\end{equation}
where $\kappa=0.5Mt $, and \emph {M} is the mass of the UAV~\cite{jeong2017mobile}. Meanwhile, the computing energy consumption of UAV in one TS can be described as
\begin{equation}
E_{n}^c (t)=\gamma_{c}C_{n}R_{n}(t)(f_{c})^{2},
\end{equation}
where $C_{n}$ expresses the number of CPU cycles needed for one bit computing, $R_{n}(t)$ represents the overall achievable sum uploading data rate of \emph{n}-th UAV in one TS, $f_{c}$ is the CPU frequency, and $\gamma_{c}$ denotes the effective switching capacitance.

\begin{figure*}[ht]
\centerline{\includegraphics[scale=0.42]{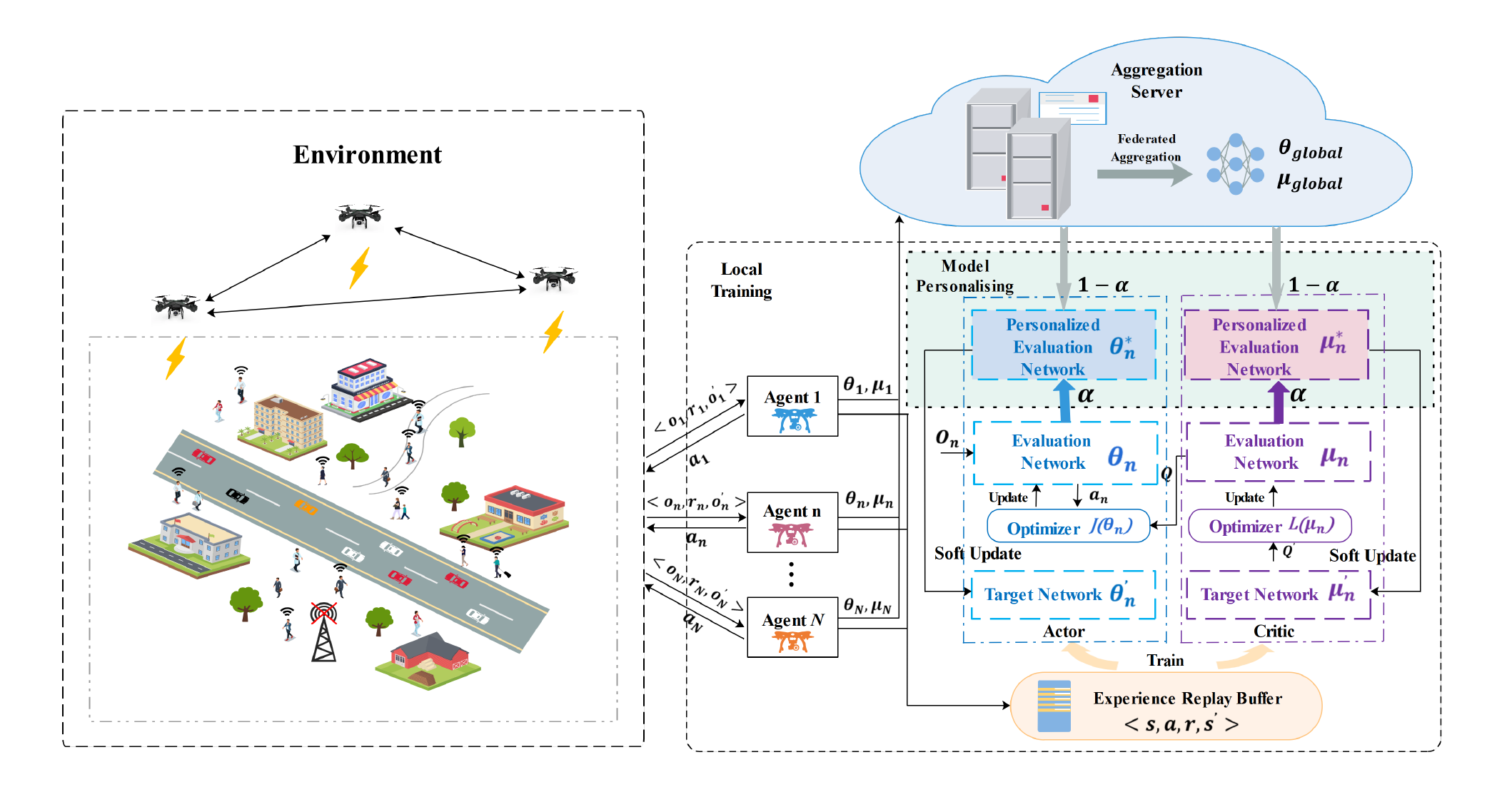}}
\caption{System model and learning framework of PF-MADDPG}
\label{fig}
\end{figure*}

The model aims to optimize the trajectory of the UAVs to maximize the quality of service. Thus, the trajectory optimization problem of multi-UAV-assisted edge computing in the dynamic environment can be formulated as
\begin{subequations}
\begin{alignat}{4}
&\max\quad \frac{\sum\limits_{n=0}\limits^{N}\sum\limits_{t=0}\limits^{T}R_{n}(t)}{NT}, \tag{6a} \vspace{2ex} \\
\mbox{s.t.}\quad
&\parallel\mathcal{P}_{i}(t)-\mathcal{P}_{j}(t)\parallel \geq \psi,  \quad&\forall i,j \in \mathcal{N}, \quad\forall t \in \mathcal{T} \tag{6b} \vspace{1ex}  \\
&\parallel x_{n}(t)\parallel \leq L^{max},  \quad&\forall n \in \mathcal{N}, \quad\forall t \in \mathcal{T} \tag{6c} \vspace{1ex} \\
&\parallel y_{n}(t)\parallel \leq L^{max},  \quad&\forall n \in \mathcal{N}, \quad\forall t \in \mathcal{T} \tag{6d} \vspace{1ex} \\
&\parallel\mathcal{P}_{i}(t)-\mathcal{P}_{b}\parallel \geq \varphi,  \quad&\forall n \in \mathcal{N}, \quad\forall t \in \mathcal{T} \tag{6e} \vspace{1ex}
\end{alignat}
\end{subequations}
where (6b) represents the minimum distance between UAVs, (6c)-(6d) states UAVs should fly within the specified range, $\mathcal{P}_{b}$ denotes the position of the obstacles and (6e) explains UAVs should keep a safe distance from obstacles. Our objective is to maximize the real-time information transmission rate between users and UAVs. Note that this problem is non-trivial to solve with traditional optimization methods since it involves continuous state space and action space, and the number of variables is unprecedented. Thus, in this paper, we propose a PF-DRL solution based on the DRL algorithm, which can achieve a fast convergence rate in a complex dynamic environment.

\section{PF-MADDPG FOR TRAJECTORY OPTIMIZATION PROBLEM}
In this section, we introduce a multi-UAV trajectory optimization scheme based on the PF-MADDPG algorithm. We first use the MADDPG algorithm to train local models. Then, the global model is trained by aggregating local models to improve the convergence rate of the algorithm. Finally, by improving the aggregation process, a personalized training solution is proposed to solve the problem that a single global model may affect the performance of local agents.

\subsection{Problem Transformation}

We first formulate the multi-UAV trajectory optimization problem as a multi-agent Markov decision process (MDP). The observation, action and reward function for each agent in \emph{t}-th TS are defined as follows:

\emph{1)} \emph{Observation $o_{n}(t)$}: The uploading data rate is closely related to the distance between the UAV and the user. All UAVs also need to obtain each other's location information. Thus, each UAV is equipped with a position acquisition device, such as a camera, so that all UAVs can obtain the position information of other UAVs and all users in real time. In summary, the observation space $o_{n}(t)$ can be expressed as: $o_{n}(t)=\{{\mathcal{P}_{n}(t), \mathcal{P}_{-n}(t),\mathcal{U}_{m, M}(t)}\}$, where $\mathcal{P}_{n}(t)$ denotes the position of \emph{n}-th UAV, $\mathcal{P}_{-n}(t)$ denotes the position of other UAVs except \emph{n}-th UAV, and $\mathcal{U}_{m, M}(t)$ denotes the position of all users.

\emph{2)} \emph{Action $a_{n}(t)$}: Each UAV needs to make action decisions in real time according to the observed environment. The action can be written as $a_{n}(t)=(\Delta x_{n}(t), \Delta y_{n}(t), \Delta z_{n}(t))$ in the TS t, where $\Delta z_{n}(t)=0$ since the UAV flies at a fixed altitude.

\emph{3)} \emph{Reward function $r_{n}(t)$}: we define the reward function as:
\begin{equation}
r_{n}(t)= \frac{1}{N}\sum_{n=1}^{N}\left({\frac{R_{n}(t)}{R_{max}} - \frac{1}{e_{max}}\left(E_{n}^f (t)+E_{n}^c (t)\right)}\right)-\varepsilon,
\end{equation}
where $R_{max}$ represents the maximum achievable rate, $e_{max}$ represents the maximum energy consumption in one time slot, and $\varepsilon$ denotes the penalty for various collisions and exceeding the boundary.

In the setting of the reward function, on the one hand, we should ensure the maximization of service quality. We should also ensure that the energy consumption of the UAVs is reduced as much as possible, so the energy consumption should have a negative impact on the reward function. Average return value in the random policy is given by $R= \frac{1}{T_0}\sum_{t=1}^{T_0}{r_{n}(t)}$, where $T_0$ denotes the total time step of a training episode.

\subsection{PF-MADDPG based Solution}
\emph{1)} \emph{Local Training}: We use the MADDPG algorithm to train the local model. MADDPG algorithm includes two main components, actor network, and critic network. The actor network outputs actions according to the policy $\pi_{n}$, and the critic network evaluates the action by calculating Q-value. In order to improve the stability of the algorithm, two target networks with the experience replay buffer, i.e., DQN, are further inserted.

The global state of the environment can be combined by the observation information of each UAV, expressed as $s(t)=\{o_{n}(t), \forall n\in \mathcal{N}\}$. Each UAV can acquire each other's observation information by communicating with each other, so all UAVs will acquire the global state $s$ and the next global state $s'$ after one action. Then, the state transition sample $\{s, a_{1}(t), a_{2}(t), ..., a_{N}(t), r_{1}(t), r_{2}(t), ..., r_{N}(t), s'\}$ is illustrated and the sample is stored in the experience replay buffer.

\begin{algorithm}[t]
    \caption{PF-MADDPG based Trajectory Optimization Algorithm}
    \begin{algorithmic}[1]
        \State Initialize: the position of UAVs $\mathcal{P}_{\emph {n }}$, users $\mathcal{U}_{\emph {m }}$;
        \State Initialize: the parameters of each UAV's actor and critic evaluation and target networks;
        \State Initialize: each UAV’s experience replay buffer;
        \For{Episode=1,2,...,$N_{max}$}
        \State Initialize the environment state $s_{0}$;
            \For{Time step=1,2,...,$\emph{T}$}
                \For{each UAV $\emph{n}$}
                \State Obtain observation $o_{n}(t)$;
                \State Select action $a_{n}(t)$ based on policy $\pi(o_{m}|\pi_{m})$;
                \State Take action $a_{n}(t)$, obtain $r_{n}(t)$;
                \EndFor
            \State obtain $s$ and $s^{'}$ by mutual communication;
                \For{each UAV $\emph{n}$}
                \State Select $a^{all}(t)$ and $r^{all}(t)$;
                \State Store sample $\{s, a^{all}(t), r^{all}(t), s^{'}\}$ into
                \State experience replay buffer;
                \State Randomly select k samples;
                \State Update critic evaluation network by (9)-(10);
                \State Update actor evaluation network by (11);
                \EndFor
            \State Aggregate the model parameters of the actor and
            \State critic evaluation network of each UAV:
            \State $\theta_{\rm global} \gets \sum_{n}\rho_{n}\theta_{n}$;
            \State $\mu_{\rm global} \gets \sum_{n}\rho_{n}\mu_{n}$;
                \For{each UAV $\emph{n}$}
                \State Model personalized training:
                \State $\theta_{n}^{*} \gets \alpha\theta_{n}+(1-\alpha)\theta_{\rm global}$;
                \State $\mu_{n}^{*} \gets \alpha\mu_{n}+(1-\alpha)\mu_{\rm global}$;
                \State Update parameters of target networks:
                \State $\theta_{n}^{'} \gets \tau\theta_{n}^{*}+(1-\tau)\theta_{n}^{'}$;
                \State $\mu_{n}^{'} \gets \tau\mu_{n}^{*}+(1-\tau)\mu_{n}^{'}$;
                \EndFor
            \State Update the current global status and enter $s^{'}$;
            \EndFor
        \EndFor
    \end{algorithmic}
\end{algorithm}

Next, by taking $k$ samples from the experience replay buffer, $\{s_{i}, a_{i,1}(t), ..., a_{i,N}(t), r_{i,1}(t), ..., r_{i,N}(t), s'_{i}\}$, where $i=1,2,...,k$, the actor target network outputs the optimal action $a_{i,n}$ under each state $s'_{i}$, and the critic target network calculates the target Q-value of the \emph{k} samples by
\begin{equation}
\begin{split}
Q(s_{i}, a_{i,1}(t), ... , a_{i,N}(t)\mid\mu'_{n})
=r_{i,n} \\ +\gamma Q(s'_{i}, a'_{i,1}(t), ... , a'_{i,N}(t)\mid\mu'_{n}).
\end{split}
\end{equation}

The parameters of the critic evaluation network can be updated by minimizing the loss function:
\begin{equation}
\begin{split}
L(\mu_{n})=\frac{1}{k}\sum_{i=1}^{k}[Q(s_{i}, a_{i,1}(t), ... , a_{i,N}(t)\mid\mu'_{n})\\-Q(s_{i}, a_{i,1}(t), ... , a_{i,N}(t)\mid\mu_{n})]^{2};
\end{split}
\end{equation}
\begin{equation}
\mu_{n}=\mu_{n}-\omega\nabla_{\mu_{n}}L(\mu_{n}),
\end{equation}
where $\omega$ represents the update step. Then, the parameters of the actor evaluation network can be updated by the policy gradient as:
\begin{equation}
\nabla_{\theta_{n}}J(\theta_{n})=\frac{1}{k}\sum_{i=1}^{k}\nabla_{a_{n}}Q(\cdot)\nabla_{\theta_{n}}\pi(o_{i,n}|\theta_{n})|_{a_{n}=\pi(o_{i,n}|\theta_{n})}.
\end{equation}

The parameters of the actor and critic target networks can be softly updated as follows~\cite{2015Continuous}:
\begin{equation}
\theta_{n}^{'}=\tau\theta_{n}+(1-\tau)\theta_{n}^{'};
\end{equation}
\begin{equation}
\mu_{n}^{'}=\tau\mu_{n}+(1-\tau)\mu_{n}^{'},
\end{equation}
where $\tau$ represents the update coefficient, generally taking a smaller value.

\emph{2)} \emph{Federated Aggregation Process}: We set up an aggregation cloud server on the ground. Each agent uploads its local model after a round of training, and the aggregation server collects all the models and obtains the global model through federation averaging, and then distributes the global model to each agent. Thus, by combining the non-independent identically distributed data of agents, the sample utilization is improved.

After the actor and critic evaluation network are trained, the network parameters are updated. Since the target network parameters are obtained from the evaluation network parameters training, each agent only needs to upload the local actor and critic evaluation network parameters. The aggregation server calculates the global parameters after collecting the local parameters of each agent. The global parameters of the actor and critic evaluation network can be calculated as:
\begin{equation}
\theta_{\rm global}=\sum_{n}\rho_{n}\theta_{n};
\end{equation}
\begin{equation}
\mu_{\rm global}=\sum_{n}\rho_{n}\mu_{n},
\end{equation}
where $\rho_{n}$ represents the weight of the \emph{n}-th agent network, and $\sum_{n}\rho_{n}=1$. Each agent updates the parameters of its target network after obtaining the global parameters.

\emph{3)} \emph{Model Personalization}: FL can effectively improve learning efficiency, however, it is worth noting that a single global model cannot be well generalized to local agents because of the heterogeneity of various data distributions. Ideally, each agent can use the global model to supplement the problem of fewer local training datasets and alleviate the negative impact of lacking personalization. Thus, the PF-MADDPG algorithm is proposed to solve this problem. In our algorithm, taking the actor evaluation network as an example, the optimization goal of each agent can be described as:
\begin{equation}
\min L(\alpha \theta_{n}+\left(1-\alpha\right) \theta_{\rm global}),
\end{equation}
where $\alpha$ is mixed weight, and $\alpha \theta_{n}+\left(1-\alpha\right) \theta_{\rm global}$ is a convex combination of the local model and the global model, namely, the personalized model. So we need to train the personalized model to update the two evaluation networks.

The architecture of the PF-MADDPG algorithm is shown in Fig. 1, and each UAV trains a local model based on its observed environment and uploads the local model to the aggregator. After obtaining the model parameters of the global actor and critic evaluation network, the global and local models are aggregated according to a certain weight ratio $\alpha$ to obtain the personalized model. After receiving the global model, the server will aggregate it with the local model as:
\begin{equation}
\theta_{n}^{*}=\alpha\theta_{n}+(1-\alpha)\theta_{\rm global};
\end{equation}
\begin{equation}
\mu_{n}^{*}=\alpha\mu_{n}+(1-\alpha)\mu_{\rm global},
\end{equation}
where $\theta_{n}^{*}$ and $\mu_{n}^{*}$ are the personalized actor and critic evaluation model parameters. The size of the mixed weight $\alpha$ will directly affect the performance of the personalized network model, which will be analyzed in the simulation section. Finally, $\theta_{n}^{*}$ and $\mu_{n}^{*}$ are used for updating the target network parameters. We provide a pseudo-code of the algorithm in Algorithm 1.

\section{PERFORMANCE EVALUATION}
In this section, we will comprehensively evaluate the learning performance of different algorithms. For simulations, we consider a 200 m $\times$ 200 m square area which consists of the random distribution of several dynamic users. Four UAVs serve users in this area, distributed in the four top corners of the square area. A safe distance of 10m shall be kept between UAVs and obstacles. The maximum speed of a UAV is set to 10 m/s, and the maximum speed of users is set to 2 m/s. There are 200 time slots in each episode of training. The average of each ten episodes of training is recorded as the return value of one training episode.

\begin{figure}[t]
\centering
    \begin{minipage}{0.48\linewidth}
        \includegraphics[width=1.7in,height=1.3in]{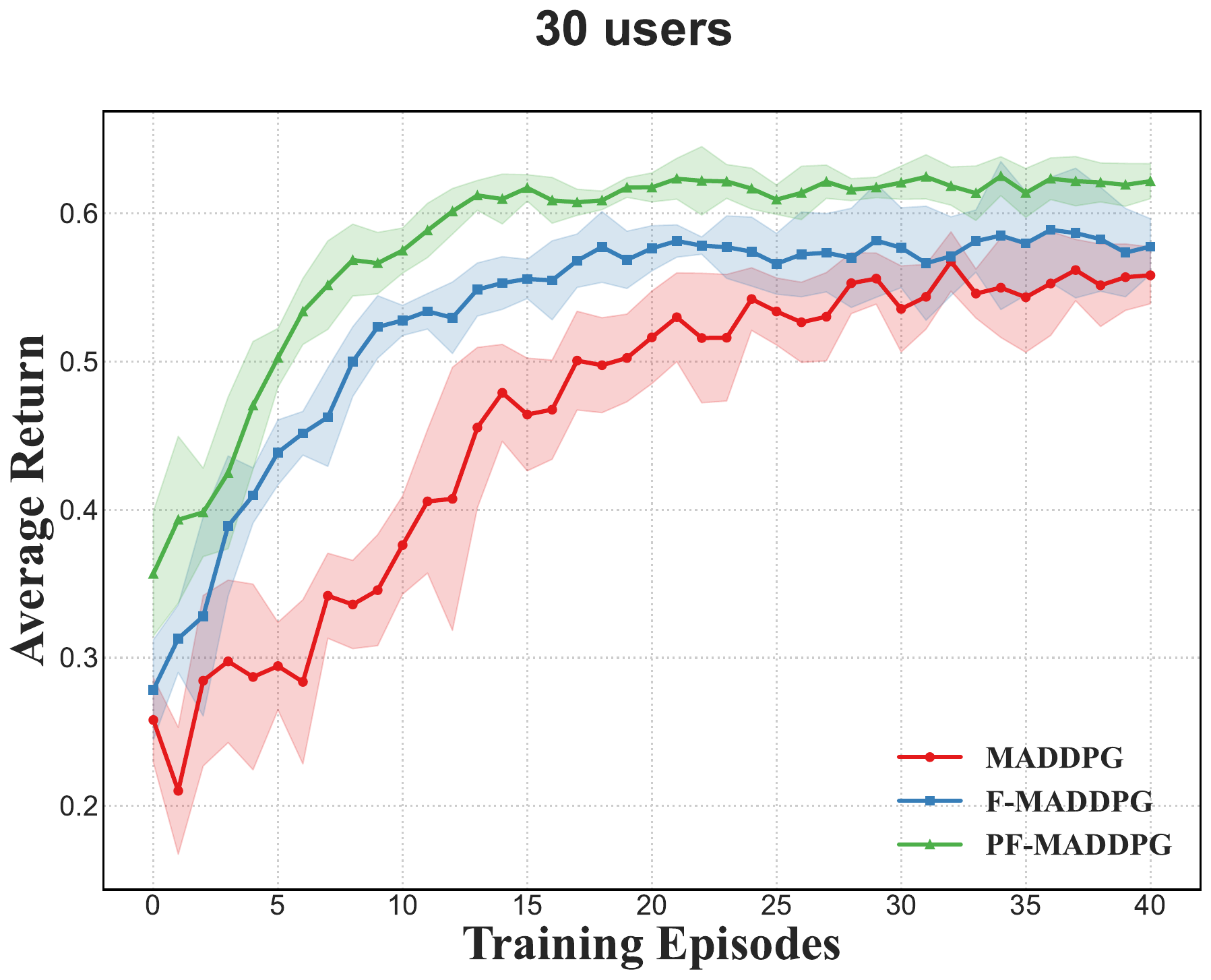}
    \end{minipage}
    \begin{minipage}{0.48\linewidth}
        \includegraphics[width=1.7in,height=1.3in]{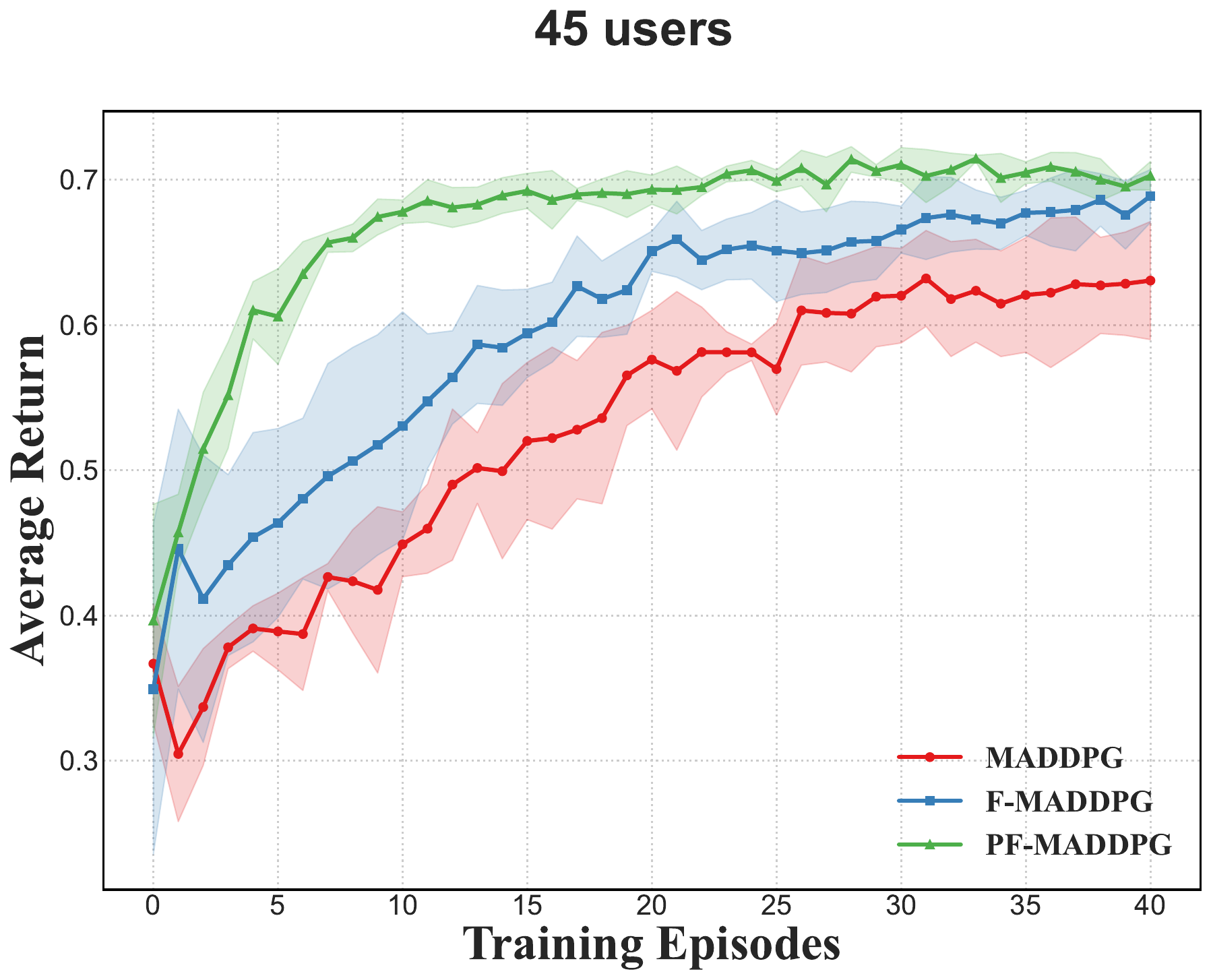}
    \end{minipage}
\centering
\caption{Performance comparison of different algorithms for 30 users and 45 users scenarios. Compared with the MADDPG algorithm and the F-MADDPG algorithm, the PF-MADDPG algorithm has a faster convergence rate and better average return.}
\end{figure}

\subsection{Overall Performance Comparison}
Fig. 2 shows the convergence of the MADDPG algorithm~\cite{wang2020multi}, the F-MADDPG algorithm, and the PF-MADDPG algorithm to optimize the trajectory of 4 UAVs, and ten rounds of experiments are conducted respectively. Four UAVs jointly serve several ground users, and each UAV can simultaneously serve five users, who share the bandwidth of the UAVs.The solid line shows the average results of the ten rounds of experiments, and the shaded part shows the range of fluctuations. 

\begin{table}[h]
\caption{The performance gain}
\centering
\begin{tabular}{@{}cccc@{}}
\toprule
\multicolumn{2}{c|}{\multirow{2}{*}{\textbf{}}}                                                                                             & \multicolumn{2}{c}{\textbf{PF-MADDPG}}                                             \\ \cmidrule(l){3-4}
\multicolumn{2}{c|}{}                                                                                                                       & \multicolumn{1}{c|}{\textbf{Average Return}} & \textbf{Convergence Rate} \\ \midrule
\multicolumn{1}{c|}{\multirow{2}{*}{\textbf{\begin{tabular}[c]{@{}c@{}}30\\ users\end{tabular}}}} & \multicolumn{1}{c|}{\textbf{MADDPG}}   & \multicolumn{1}{c|}{\textbf{14.5\%}}              & \textbf{115.4\%}               \\ \cmidrule(l){2-4}
\multicolumn{1}{c|}{}                                                                             & \multicolumn{1}{c|}{\textbf{F-MADDPG}} & \multicolumn{1}{c|}{\textbf{10.5\%}}              & \textbf{38.5\%}                \\ \midrule
\multicolumn{1}{c|}{\multirow{2}{*}{\textbf{\begin{tabular}[c]{@{}c@{}}45\\ users\end{tabular}}}} & \multicolumn{1}{c|}{\textbf{MADDPG}}   & \multicolumn{1}{c|}{\textbf{14.3\%}}              & \textbf{136.4\%}               \\ \cmidrule(l){2-4}
\multicolumn{1}{c|}{}                                                                             & \multicolumn{1}{c|}{\textbf{F-MADDPG}} & \multicolumn{1}{c|}{\textbf{7.4\%}}              & \textbf{81.1\%}                \\ \bottomrule
\end{tabular}
\end{table}

The simulation results indicate that the PF-MADDPG algorithm has a faster convergence rate than the MADDPG algorithm and the F-MADDPG algorithm. Conversely, the training process of the MADDPG algorithm and the MADPPG algorithm is exceedingly unstable and subject to considerable fluctuations when compared to the PF-MADDPG algorithm. TABLE 1 provides a comprehensive overview of the performance and convergence rate gain of the PF-MADDPG algorithm across two distinct environments. In general, the PF-MADDPG algorithm showcases superior average return and faster convergence speed, coupled with notable performance improvements.

\subsection{Local Performance Analysis}
The simulation analysis above indicates that the PF-MADDPG algorithm significantly improves the overall convergence rate of the system. However, the benefits of personalized federated learning extend beyond just the convergence rate, as it also enhances the local performance of each agent through personalized model training. Thus, in this subsection, we analyze the local performance of a single agent. Fig. 3 illustrates weighted statistics of the local return of each agent, comparing the PF-MADDPG algorithm with the MADDPG algorithm. The simulation analysis shows that the PF-MADDPG algorithm yields a performance gain of up to $13\%$-$18\%$ for each UAV, resulting in substantial improvement in local performance compared to the MADDPG algorithm.

\begin{figure}[h]
\centerline{\includegraphics[scale=0.435]{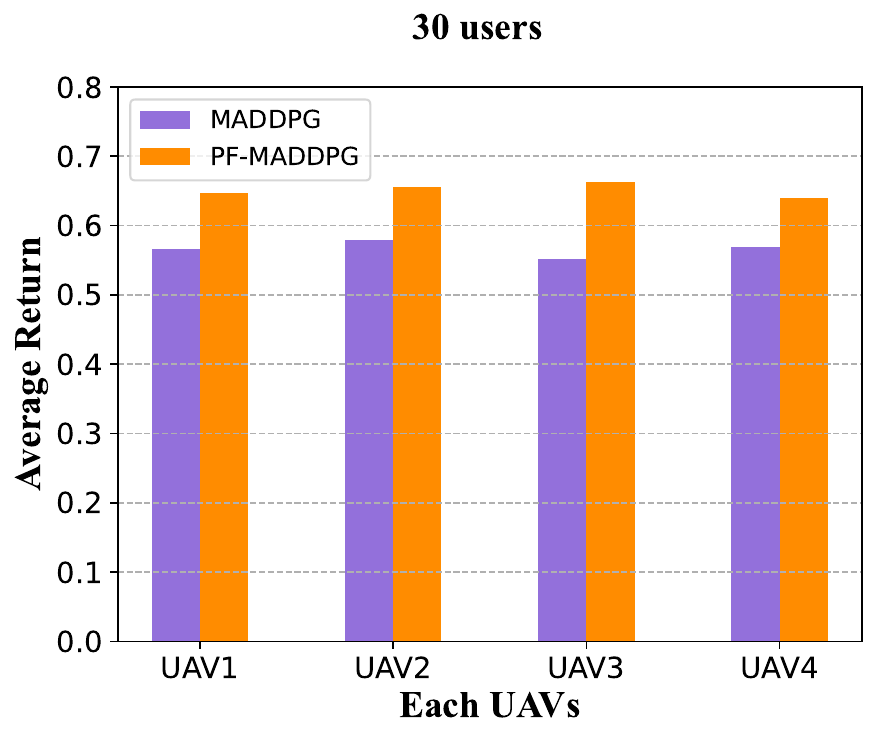}}
\caption{Comparison of local performance between PF-MADDPG algorithm and MADDPG algorithm in 30 users scenario.}
\label{fig}
\end{figure}

\subsection{Discussion on the Mixed Weight $\alpha$}
In this subsection, we analyze the weight of the global model in the PF-MADDPG algorithm, that is, the aggregate ratio of the local model to the global model. In the experiment, we find that different weight ratios will affect the overall performance of the personalized model. As shown in Fig. 4, we have set several fixed weight ratios. The weight ratios of the private model and the aggregate model are 3:7, 5:5, 7:3, and 9:1 respectively for the simulation experiments, and are compared with the MADDPG algorithm. From the experimental results, it can be concluded that in the UAV trajectory optimization problem, the private model should have more weight, and the optimal ratio should be around 7:3. However, it is difficult to find the best-mixed weight. An optimized $\alpha$ is decided by a combination of several system factors, such as the correlation between a local distribution and global distribution\cite{deng2020adaptive}, which can be further explored in the future.

\begin{figure}[h]
\centerline{\includegraphics[scale=0.215]{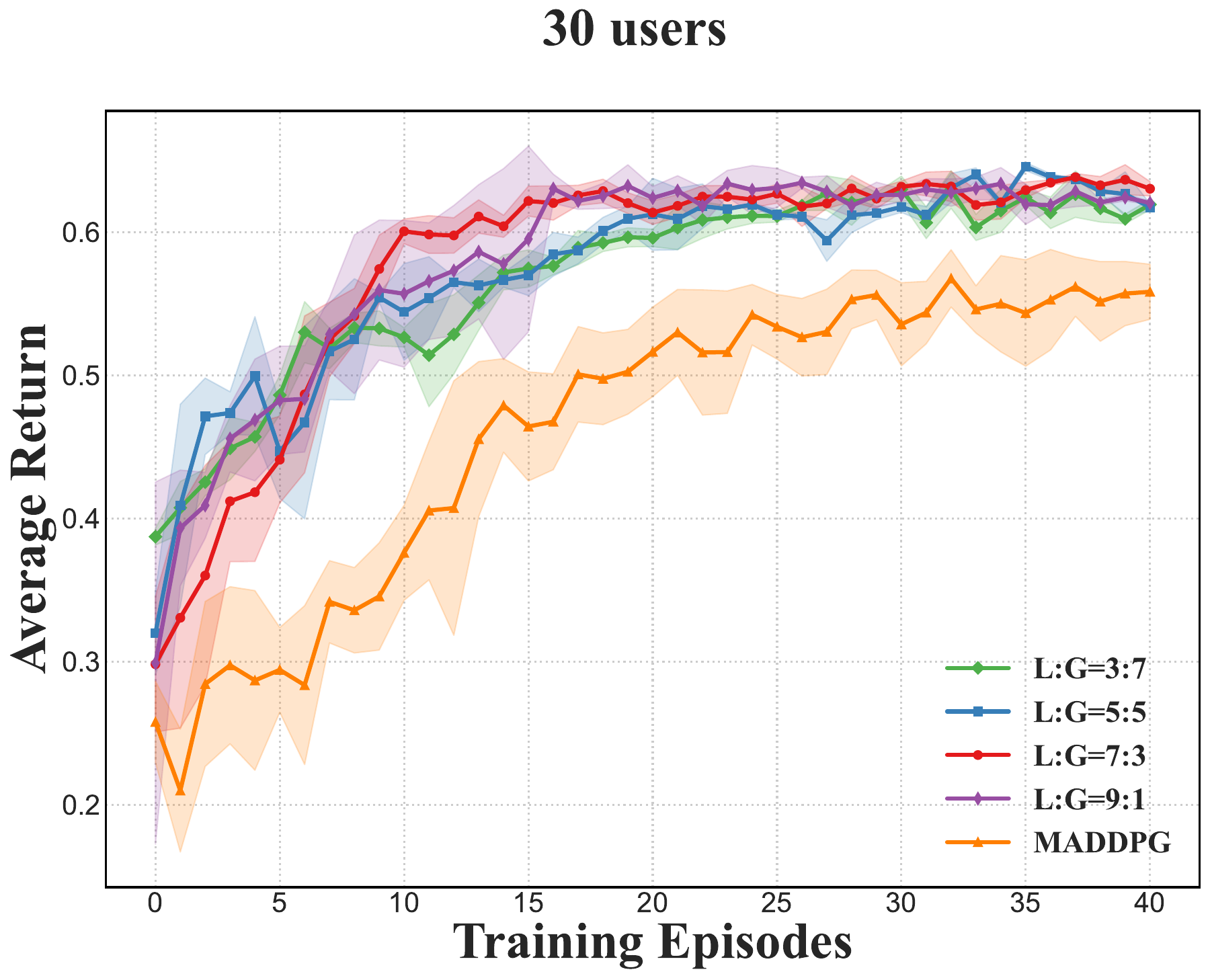}}
\caption{Performance comparison of personalized federated reinforcement learning algorithms with different mixed weights. The mixture weight ratio shown in the figure is the ratio of the local model to the global model(L: G).}
\label{fig}
\end{figure}

\section{CONCLUSION}
In this paper, we have proposed a personalized federated deep reinforcement learning algorithm to solve the trajectory optimization problem of multi-UAV in complex dynamic scenarios. By aggregating local and global models with a certain mixture weight, a well-trained personalized model can be obtained by several rounds of communication. Simulation results show that the proposed scheme can achieve a faster convergence rate and better learning performance, compared with the MADDPG algorithm and the F-MADDPG algorithm.

\bibliographystyle{IEEEtran}

\begin{thebibliography}{10}
\providecommand{\url}[1]{#1}
\csname url@samestyle\endcsname
\providecommand{\newblock}{\relax}
\providecommand{\bibinfo}[2]{#2}
\providecommand{\BIBentrySTDinterwordspacing}{\spaceskip=0pt\relax}
\providecommand{\BIBentryALTinterwordstretchfactor}{4}
\providecommand{\BIBentryALTinterwordspacing}{\spaceskip=\fontdimen2\font plus
\BIBentryALTinterwordstretchfactor\fontdimen3\font minus
  \fontdimen4\font\relax}
\providecommand{\BIBforeignlanguage}[2]{{%
\expandafter\ifx\csname l@#1\endcsname\relax
\typeout{** WARNING: IEEEtran.bst: No hyphenation pattern has been}%
\typeout{** loaded for the language `#1'. Using the pattern for}%
\typeout{** the default language instead.}%
\else
\language=\csname l@#1\endcsname
\fi
#2}}
\providecommand{\BIBdecl}{\relax}
\BIBdecl

\bibitem{corcoran2016mobile}
P.~Corcoran and S.~K. Datta, ``{Mobile-edge computing and the {Internet of
  Things} for consumers: Extending cloud computing and services to the edge of
  the network},'' \emph{IEEE Consumer Electronics Magazine}, vol.~5, no.~4, pp.
  73--74, 2016.

\bibitem{fotouhi2019survey}
A.~Fotouhi, H.~Qiang, M.~Ding, M.~Hassan, L.~G. Giordano, A.~Garcia-Rodriguez,
  and J.~Yuan, ``{Survey on UAV cellular communications: Practical aspects,
  standardization advancements, regulation, and security challenges},''
  \emph{IEEE Communications Surveys \& Tutorials}, vol.~21, no.~4, pp.
  3417--3442, 2019.

\bibitem{1}
Y.~Zeng and R.~Zhang, ``{Energy-efficient UAV communication with trajectory
  optimization},'' \emph{IEEE Transactions on Wireless Communications},
  vol.~16, no.~6, pp. 3747--3760, 2017.

\bibitem{9727746}
M.~Guo, T.~Long, H.~Li, and J.~Sun, ``{Reinforcement-learning-based path
  planning for UAVs in intensive obstacle environment},'' in \emph{2021 China
  Automation Congress (CAC)}, 2021, pp. 6451--6455.

\bibitem{yan2018path}
C.~Yan and X.~Xiang, ``{A path planning algorithm for UAV based on improved
  Q-learning},'' in \emph{2018 2nd International Conference on Robotics and
  Automation Sciences (ICRAS)}.\hskip 1em plus 0.5em minus 0.4em\relax IEEE,
  2018, pp. 1--5.

\bibitem{9739975}
Y.~Qian, K.~Sheng, C.~Ma, J.~Li, M.~Ding, and M.~Hassan, ``{Path planning for
  the dynamic UAV-aided wireless systems using Monte Carlo Tree Search},''
  \emph{IEEE Transactions on Vehicular Technology}, vol.~71, no.~6, pp.
  6716--6721, 2022.

\bibitem{nguyen2020deep}
T.~T. Nguyen, N.~D. Nguyen, and S.~Nahavandi, ``{Deep reinforcement learning
  for multiagent systems: A review of challenges, solutions, and
  applications},'' \emph{IEEE Transactions on Cybernetics}, vol.~50, no.~9, pp.
  3826--3839, 2020.

\bibitem{9465671}
P.~Luong, F.~Gagnon, L.-N. Tran, and F.~Labeau, ``{Deep reinforcement
  learning-based resource allocation in cooperative UAV-assisted wireless
  networks},'' \emph{IEEE Transactions on Wireless Communications}, vol.~20,
  no.~11, pp. 7610--7625, 2021.

\bibitem{bayerlein2021multi}
H.~Bayerlein, M.~Theile, M.~Caccamo, and D.~Gesbert, ``{Multi-UAV path planning
  for wireless data harvesting with deep reinforcement learning},'' \emph{IEEE
  Open Journal of the Communications Society}, vol.~2, pp. 1171--1187, 2021.

\bibitem{wang2020multi}
L.~Wang, K.~Wang, C.~Pan, W.~Xu, N.~Aslam, and L.~Hanzo, ``{Multi-agent deep
  reinforcement learning-based trajectory planning for multi-UAV assisted
  mobile edge computing},'' \emph{IEEE Transactions on Cognitive Communications
  and Networking}, vol.~7, no.~1, pp. 73--84, 2020.

\bibitem{zhao2022multi}
N.~Zhao, Z.~Ye, Y.~Pei, Y.-C. Liang, and D.~Niyato, ``{Multi-agent deep
  reinforcement learning for task offloading in UAV-assisted mobile edge
  computing},'' \emph{IEEE Transactions on Wireless Communications}, 2022.

\bibitem{gao2021game}
A.~Gao, Q.~Wang, W.~Liang, and Z.~Ding, ``{Game combined multi-agent
  reinforcement learning approach for UAV assisted offloading},'' \emph{IEEE
  Transactions on Vehicular Technology}, vol.~70, no.~12, pp. 12\,888--12\,901,
  2021.

\bibitem{2021A}
Q.~Xia, W.~Ye, Z.~Tao, J.~Wu, and Q.~Li, ``{A survey of federated learning for
  edge computing: research problems and solutions},'' \emph{High-Confidence
  Computing}, 2021.

\bibitem{kwon2020multiagent}
D.~Kwon, J.~Jeon, S.~Park, J.~Kim, and S.~Cho, ``{Multiagent DDPG-based deep
  learning for smart ocean federated learning IoT networks},'' \emph{IEEE
  Internet of Things Journal}, vol.~7, no.~10, pp. 9895--9903, 2020.

\bibitem{ziying2021towards}
A.~Ziying~Tan, H.~Yu, L.~Cui, and Q.~Yang, ``Towards personalized federated
  learning,'' \emph{arXiv e-prints}, pp. arXiv--2103, 2021.

\bibitem{jeong2017mobile}
S.~Jeong, O.~Simeone, and J.~Kang, ``{Mobile edge computing via a UAV-mounted
  cloudlet: Optimization of bit allocation and path planning},'' \emph{IEEE
  Transactions on Vehicular Technology}, vol.~67, no.~3, pp. 2049--2063, 2017.

\bibitem{2015Continuous}
T.~P. Lillicrap, J.~J. Hunt, A.~Pritzel, N.~Heess, T.~Erez, Y.~Tassa,
  D.~Silver, and D.~Wierstra, ``{Continuous control with deep reinforcement
  learning},'' \emph{Computer ence}, 2015.


\bibitem{deng2020adaptive}
Y.~Deng, M.~M. Kamani, and M.~Mahdavi, ``{Adaptive personalized federated
  learning},'' \emph{arXiv preprint arXiv:2003.13461}, 2020.







\end{thebibliography}

\end{document}